# Discovering Equations that Govern Experimental Materials Stability under Environmental Stress using Scientific Machine Learning

*Richa Ramesh Naik\*, Armi Tiihonen\*, Janak Thapa, Clio Batali, Zhe Liu, Shijing Sun, Tonio Buonassisi\**

Massachusetts Institute of Technology, 77 Massachusetts Avenue, Cambridge, MA 02139

\*Correspondence to: RN (richarnaik96@gmail.com), AT (armi.tiihonen@gmail.com), TB (buonassisi@mit.edu)

*June 21, 2021 version*

**Abstract:**

While machine learning (ML) in experimental research has demonstrated impressive predictive capabilities, inductive reasoning and knowledge extraction remain elusive tasks, in part because of the difficulty extracting fungible knowledge representations from experimental data. In this manuscript, we use ML to infer the underlying dynamical differential equation (DE) from experimental data of degrading organic-inorganic methylammonium lead iodide (MAPI) perovskite thin films under environmental stressors (elevated temperature, humidity, and light). We apply a sparse regression algorithm that automatically identifies the differential equation describing the dynamics from time-series data. We find that the underlying DE governing MAPI degradation across a broad temperature range of 35 to 85°C is described minimally with three terms (specifically, a second-order polynomial), and not a simple single-order reaction (*i.e.* 0[th], 1[st], or 2[nd]-order reaction). We demonstrate how computer-derived results can aid the researcher to develop profound mechanistic insights. This DE corresponds to the Verhulst logistic function, which describes reaction kinetics analogous in functional form to autocatalytic or self-propagating reactions, suggesting future strategies to suppress MAPI degradation. We examine the robustness of our conclusions to experimental luck-of-the-draw variance and Gaussian noise using a combination of experiment and simulation, and describe the experimental limits within which this methodology can be applied. Our study demonstrates the application of scientific ML in experimental chemical and materials systems, highlighting the promise and challenges associated with ML-aided scientific discovery.

**Introduction**

In the traditional scientific discovery process, prior knowledge from first-principles and empirical laws are combined with experimental data and intuition to yield governing equations. Newton's law of gravitation[1], Einstein's mass-energy equivalence equation[2], Kepler's laws of planetary motion[3] and other physical principles were uncovered through careful interpretation of experimental data and inductive reasoning[4]. The approach of fitting experimental data through regression is difficult with systems that are yet to be understood fully – the set of feasible equations capturing the physics is enormous.[5–7]

One such area where underlying physics is often poorly understood is the study of materials under environmental stress. For example, alloys[8,9], polymers[10], doped silicon[11] and hybrid materials[12] experience changes at elevated temperatures. The degradation pathways can be complex and not directly obvious when examining the experimental data. Machine learning (ML) has been used to predict degradation[13–17] as well as to optimize process conditions to reduce material decomposition[16,18]. However, traditional data-science methods yield little insight into the underlying mechanisms. We posit that hidden in the black-box ML models is valuable scientific information on the dynamics of the system. If uncovered, the knowledge of the governing dynamics can serve as foundation for physical interpretation of phenomena and scientific discovery.

Herein, we use scientific ML, which combines regression-based ML with sparsity generating techniques in order to automatically identify governing equations directly from data, especially when the systems being studied are too complicated to yield to traditional theoretical analysis. Not only does scientific ML help us understand the underlying scientific phenomena better, it also has the potential to help to make simulations faster and extrapolate beyond the dataset at hand.

Recently, many approaches aiming for this target have been presented in literature. A method that we apply in this contribution is PDE-FIND by Rudy *et al.*[19] This method is used for the discovery of physical laws describing dynamical systems. First, a library of potential candidate functions is built. Differentials are calculated by finite difference or polynomial interpolation. Once a large matrix with all candidate functions is composed, different sparse regression methods may be used to extract the partial differential equation (PDE) describing the system. The sparse methods implemented are sequential threshold ridge regression, lasso regression, elastic net regression and greedy algorithm. Another sparse technique is Sparse Identification of nonlinear Dynamics (SINDy)[20]. It uses a custom deep autoencoder to find a coordinate system in which the dynamics of the system are sparse, and then uses sparse regression to find the governing equations in the associated coordinate system. Atkinson et al.[21] present a generalized method for the discovery of differential equations using genetic programming. Physics Informed Neural Networks (PINN) [22] and PDE-NET[23,24] are deep learning methodologies to extract governing partial differential equations using dynamical data. These methods have shown great promise in several applications[25–28]. The automatic discovery of scientific laws and principles is at the frontier of machine learning that awaits application to materials science[29] and other domains[30–32].

Halide perovskite materials, which have potential to provide high performing and cost-effective solar energy, degrade at elevated temperature[33–38], humidity[39–41], and illumination[42,43]. This is a major issue hindering the commercialization of perovskite photovoltaic technology. However, the degradation mechanisms affecting halide perovskites are not well understood. Discovering the underlying equations directly from perovskite degradation data could accelerate the development of stable perovskite solar cells. Herein, we apply Scientific ML to study the environmental degradation of methylammonium lead iodide (MAPI).

From prior knowledge in the literature, MAPI has multiple documented reaction pathways, including decomposition to PbI$_2$ via reaction[44]:

$$MAPbI_3 \rightarrow PbI_2 + [CH_3NH_3^+ + I^-] \rightarrow PbI_2 + CH_3NH_2 + HI \qquad (1)$$

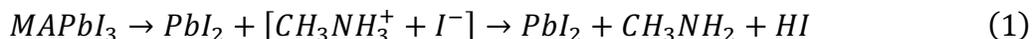

Smecca et al.[45] demonstrate that the rate of MAPI degradation obeys an Arrhenius-type law. Their data suggests that the degradation of MAPI follows zero-order kinetics in the presence of moisture and first-order kinetics in vacuum at temperatures ranging from 90°C to 135°C. Bastos et al.[46] hypothesize that the thermal degradation of MAPI is defined by the Avrami equation[47,48] of nucleation and growth. The Avrami equation has also been used to describe degradation kinetics in humid air[49]. Recently, studies have shown that halide perovskite degradation follows autocatalytic reaction kinetics[50] with the hypothesis that the degradation is propagated by iodine vapors[51]. The derivation of exact kinetics through first principles as well as Arrhenius-type dependence is difficult because of the complexity of MAPI decomposition, despite the availability of well-resolved dynamical data, inviting the application of Scientific ML.

In this study, we focus on the application of PDE-FIND to perovskite degradation data. We choose PDE-FIND as it is an interpretable method that provides a parsimonious description of the dynamics with the flexibility to apply domain expertise for library selection. Successfully identifying governing differential equations directly from the experimental aging test data would deepen the understanding of thermal degradation and provide tools for reliable lifetime prediction of perovskite solar cells as well as the determination of acceleration factors for long-term aging tests. These developments could spur the advancement of the perovskite photovoltaic technology and have been called for by the community[52–54]. This study provides a generalizable pathway to identify degradation modes in other materials research domains as well.

**Methods**

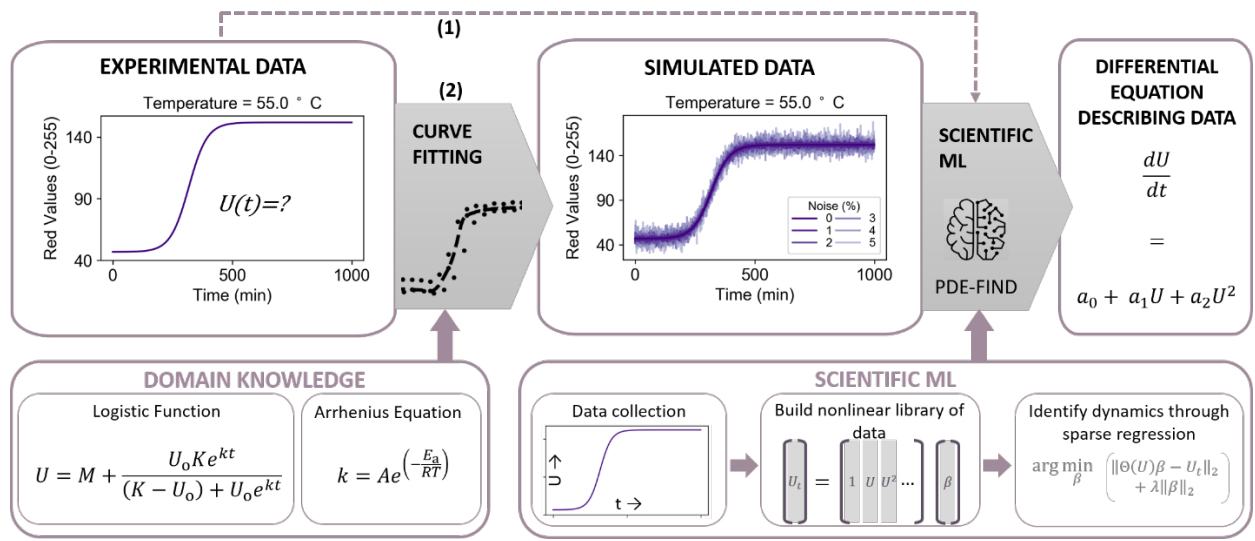

**Figure 1.** Schematic of the data-management workflow used in this study. Workflow (1) applies PDE-FIND directly to experimental data; Workflow (2) first fits the experimental data with a logistic function to create a simulated dataset, optionally adds Gaussian noise, and then applies PDE-FIND.

A summary of our data-analysis workflow is shown in **Figure 1**. Our goals for this study are two-fold: Uncover the underlying differential equation corresponding to perovskite degradation using sparse regression methodology PDE-FIND (Workflow **(1)**) and quantify the effect of noise on the accuracy of extraction of differential equations by PDE-FIND by comparing noiseless and noisy simulated data (Workflow **(2)**). This is represented by the two workflows in Figure 1.

For the first objective, the input is the experimental data obtained from degrading MAPI films. Our experimental data is shown in **Figure 2**. We subjected 206 thin-film samples of methylammonium lead iodide (MAPI) to 0.15±0.01 Sun illumination, 20±5% relative humidity, and temperatures varying from 35 to 85°C in our in-house environmental chamber described in detail in Ref. 18 (Figure 2b). One hundred and eight samples were grown under low-variance conditions (labeled "low-variance experimental"); ninety-eight samples were grown under high-variance conditions (labeled "high-variance experimental") (Figure S2). Unless specified otherwise, we assume "experimental" data in this paper refers to the low-variance sample set. We quantify this variance in the beginning of the "Results" section.

We monitored the degradation of MAPI based on the color change of the material. As MAPI films decompose, they change their color from initial black (majority MAPI) to degraded yellow (minority MAPI). We acquired images of the degrading films with 0.5-minute temporal resolution and processed them to obtain the average red, blue and green color components of the films as a function of time (Figure 2a, Figure S1). The red color time-series is chosen for further analysis because it sufficiently captures the temporal perovskite decomposition behavior at the MAPI bandgap, as shown in the Supporting Information of Ref. 16 (Figure 2d, Figure S1).

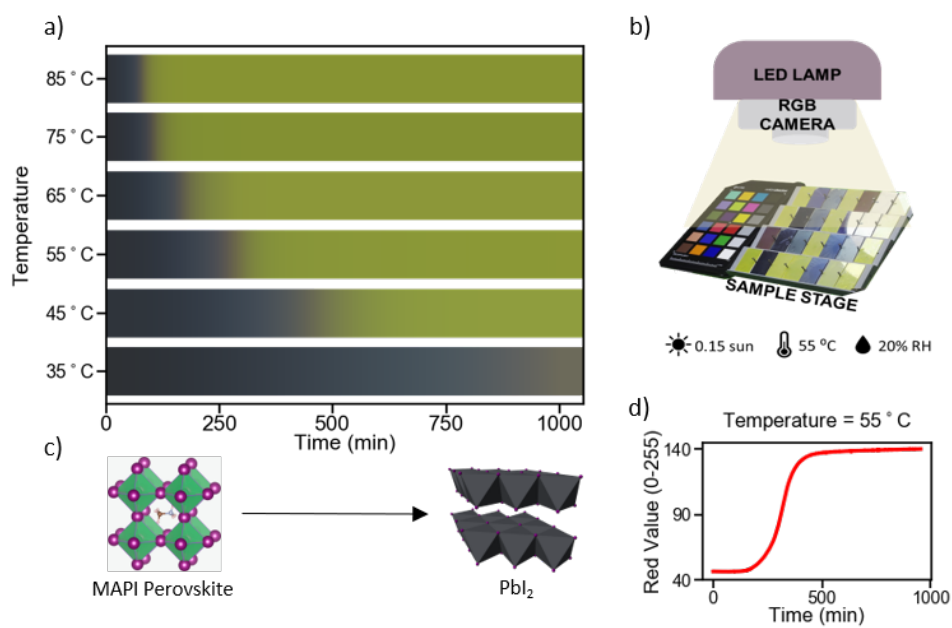

**Figure 2.** The experimental data. a) Average perovskite film color as a function of time at different temperatures. b) A schematic diagram of the in-house accelerated degradation chamber with a superimposed camera image of degrading MAPI films. c) The structure of MAPI perovskite (reactant) and lead iodide (degradation product). d) Processed average red color component of films degraded at $T = 55$ °C as a function of time.

For the second objective, we generate simulated degradation data to analyze how noise obfuscates the identification of underlying DEs. We apply a non-linear least-squares method to fit the experimental data (*e.g.*, those shown in Figure 2d) to the Verhulst logistic equation[55] to model the S-shaped curve. This is a reasonable assumption because the logistic function is used to describe the thermal decomposition dynamics of several materials[50,51,56]. We obtain,

$$U = M + \frac{U_o K e^{kt}}{(K - U_o) + U_o e^{kt}}, \quad (2)$$

$$\frac{\partial U}{\partial t} = k(U - M)\left(1 - \frac{(U - M)}{K}\right) \quad (3)$$

where $U_o$ is the initial concentration, $k$ is growth rate, $K$ is the carrying capacity and $M$ is a fitting constant. In the context of MAPI degradation, $M$, $U_o$ and $K$ can be considered as fitting parameters. The growth rate $k$ varies with temperature according to the Arrhenius equation:

$$k = A e^{\left(-\frac{E_a}{RT}\right)} \quad (4)$$

Here, $E_a$ is the activation energy, $T$ is the temperature in Kelvin, $A$ is the pre-exponential factor and $R$ is the universal gas constant. We use this model to produce noise-free simulated data (labeled "simulated") and simulated data with Gaussian noise (labelled "simulated with Gaussian noise").

First, we apply the sparse regression methodology PDE-FIND[19] to experimental data (Workflow **(1)**). We use the time-series from all the temperatures to infer the partial differential equation (PDE) defining the relationship between MAPI degradation, temperature and time. Then, we apply PDE-FIND to the time-dependent degradation data at each temperature, to infer the ordinary differential equation (ODE) that describes MAPI decomposition at a particular temperature. To study the effect of noise, we apply PDE-FIND to simulated data with and without Gaussian noise (Workflow **(2)**).

The library of potential candidate functions consists of polynomials of $U$, polynomials of time $t$, sine and cosine of $U$, Temperature $T$ and other non-linear functions of $U$, $t$ and $T$ (Table S1). Differentials are calculated by finite difference with convolutional smoothing using a 1D Gaussian kernel. Once a large tall matrix ($\Theta(U)$) with all candidate functions is composed, we use sequential threshold ridge regression to identify which terms contribute to the dynamics described by the data as well as those terms' weights. The goal of this method is to find a sparse coefficient vector $\beta$ that only consists of the active features that best represent the time derivative $U_t$. The rest of the features are hard-thresholded to zero. The loss functions are follows ($\lambda_2$ and $\lambda_0$ are the L-2 and L-0 regularization penalties respectively, more details can be found in in the supplementary information of Ref. 19):

$$\hat{\beta} = \arg\min_{\beta}( \|\Theta(U)\beta - U_t\|_2 + \lambda_2\|\beta\|_2) \quad (5.1)$$

for a given $\widehat{tol}$, where $\widehat{tol}$ is:

$$\widehat{tol} = \arg\min_{tol}( \|\Theta(U)\beta - U_t\|_2 + \lambda_0\|\beta\|_0) \quad (5.2)$$

## Results

Our aim is to obtain the equation that most accurately describes the environmental degradation of methylammonium lead iodide (MAPI) as a function of time and temperature. There are two main challenges for scientific ML in this application that are also common with many other experimental applications, especially in materials science: The function space that could in principle capture the degradation processes is enormous, complicating identification of unique equations. Furthermore, experimental data has measurement noise as well as sample-to-sample variance, making the identification of quantitative analytic descriptions even more challenging. These conditions can be optimized to some extent, but not excluded.

Our experimental setup represents a typical materials science experiment: The noise in our experimental data is of the order of 0.35% for both high-variance and low-variance experimental data sets. The low value indicates that the camera measurement of degradation is optimized. The sample-to-sample variance for the "low variance experimental" dataset is estimated to be 20% in relative standard deviation and the maximum mean absolute deviation is 12 units (Red color value varies from 0-255). For the "high variance experimental" dataset, variance is estimated to be 23% in relative standard deviation and the maximum mean absolute deviation is 31 units. These values are typical for spin-coated perovskite film samples that tend to have rather high variations, especially when aged.

Our workflow shown in Figure 2. First, we attempt to uncover the differential equation governing perovskite degradation directly from experimental data (Workflow **(1)**). A simple way to analyze reaction rate orders is to fit the data to pure $0^{th}$, $1^{st}$ and $2^{nd}$ order dynamics (Figure S3). These equations do not fit the data, showing that the environmental degradation of MAPI does not follow a simple *n*-th order kinetics. This motivates the use of PDE-FIND. We apply sparse regression to the *whole experimental dataset* with a broad function library consisting of polynomials of *U* up to order 5, sine and cosine of *U*, polynomials of *t* up to order 3, the square root of *t*, *U* multiplied with polynomials of *t*, temperature *T* and adjusted negative exponent of $1/T$ ($\exp(-\frac{100}{T})$). Sine and cosine terms are not selected by PDE-FIND –indicating as a sanity check that the algorithm correctly identifies that periodicity is not a feature of the dynamics. Polynomials of *t* and *U* times the polynomials of *t*, which correspond to the Avrami equation, are not included in the library or assigned very small weights. To understand how well the obtained DE represents our data, we compare the derivative estimated by our DE to the numerical derivative obtained from the experimental data. While certain trends in the derivative are captured, errors exist because of the variance in our experimental data (Figure S4). Refinements to the approach are thus needed.

We proceed to narrow the application of PDE-FIND, by applying PDE-FIND to the averaged data at each temperature individually to extract the governing ODE. Using the averaged data helps us deal with sample-to-sample variance. Since all environmental conditions were almost identical for samples degraded at a particular temperature but aging tests of each temperature were conducted one after another (introducing differences *e.g.* in sample storage times and exact equipment atmosphere), we aim to reduce the influence of variance-inducing conditions by applying PDE-FIND at each temperature separately. First, we apply PDE-FIND with a large library as described in the previous paragraph. Here too, we see that sine and cosine of *U*, polynomials of *t* and *U* times the polynomials of *t* are either removed from the library or have small coefficient values. We exclude these terms in further analysis. Then, we apply PDE-FIND with $1^{st}$ to $5^{th}$ order polynomial libraries. We find that with the $1^{st}$ order polynomial library, PDE-FIND is unable to find an equation that fits the derivative of our data (Figure 3a). All other libraries from $2^{nd}$ order polynomial to $5^{th}$ order polynomial appear to fit the derivative of our data with significant accuracy (Figure 3a, 3b). When these differential equations are integrated, they have the same S-shape as our experimental data (Figure 3c). The $2^{nd}$ order polynomial library is the most minimal library that fits our data without high error. The functional form of this ODE is:

$$\frac{dU}{dt} = a_0 + a_1 U + a_2 U^2 \tag{6}$$

We also notice a trend in the values of the fitting coefficients with respect to temperature – especially in the case of the $2^{nd}$ order polynomial library (Figure 3d, Figure S5). Then slope of the curve changes between 55 °C and 65 °C, the temperature at which a well-known MAPI phase transition[57,58] occurs. This may indicate that the phase transition affects the degradation mechanism, but is not experimentally confirmed in this work.

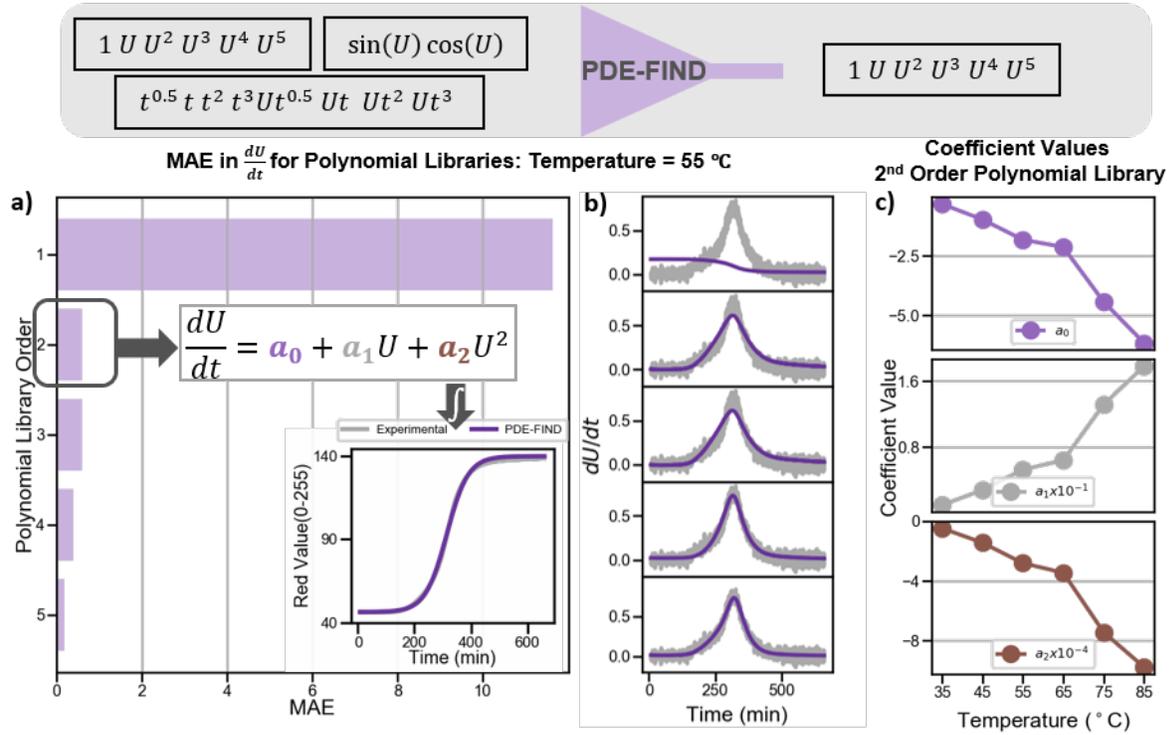

**Figure 3.** PDE-FIND results with experimental data: a) A bar plot shows the MAE between the actual experimental derivative (smoothened) and the value of the derivative estimated using the differential equation identified by PDE-FIND. Inset: Comparison of the experimental data with the curve obtained by integrating the equation identified by PDE-FIND with 2nd order polynomial library . b) Comparison of the *dU/dt* calculated from experimental data for *T* = 55 °C and estimated from PDE-FIND for 2nd order polynomial library to 5th order polynomial libraries c) Coefficient values estimated by PDE-FIND as a function of temperature for 2nd order polynomial library.

Next, we evaluate the effect of variance on PDE extraction by comparing the above results (obtained on the low-variance experimental dataset) with the same workflow applied to the high-variance data (Figure S6). After averaging multiple curves (*U*(*t*)) for each temperature, the results are qualitatively similar for a constrained function library of polynomials of 2nd order – the obtained coefficients have the same sign and order of magnitude. This indicates that PDE-FIND can fit even high-variance experimental data when appropriately averaging over multiple samples. To quantify the effect of sample-to-sample variance, we apply PDE-FIND to each curve individually. As expected, PDE-FIND extracts a large variance in coefficient values. The values of coefficients vary as much as 60% with the low variance dataset and up to 90% with the high variance datasets for T = 55 °C.

Now, we evaluate the effect of noise on PDE extraction using simulated data. We use the non-linear least-squares method to fit our experimental data to the Verhulst logistic equation[55] and the Arrhenius equation, as shown in the Methods section. We produce both noise-free simulated data and simulated data with Gaussian noise (Workflow **(2)**) with this model.

We apply sparse regression to the simulated dataset at each temperature individually to discover the governing ODEs with libraries ranging from 2nd to 5th order polynomials. With the noise-free data, PDE-FIND's identified DEs fit the derivative as well as the data on integration of the DE with significant accuracy for libraries from 2nd order to 5th order. In the case of the 2nd order polynomial library, both the underlying differential equation and the fitting parameters are identified with significant accuracy, as shown in **Figure 4**. We know that the underlying governing equation for this dataset should have terms of orders higher than 2 equals to zero – PDE-FIND assigns small non-zero values to these functional forms, although they are not set to zero. In the case where sine and cosine are added to the library, the algorithm correctly identifies that these terms do not represent the dynamics and are set to zero exactly. The MAE between the exact numerical derivative and one estimated from the differential equation identified by PDE-FIND is of order $10^{-7}$ (when derivative varies from 0 to 1). This indicates that PDE-FIND works well for simulated curves with zero noise. Thus, with the candidate function library constrained to polynomials of $U$, PDE-FIND is able to identify the same ODE that fits the data at each temperature.

We then add varying amounts of Gaussian noise to this simulated equation at different temperatures. First, we consider the effect of varying amounts of noise at a fixed temperature of 55°C, as indicated by the black box in Figure 4a. We add up to 5% noise, which is typical in many experimental settings. The equation identified by PDE-FIND yields an S-shaped curve similar to the noise-free simulated curve upon integration (Figure 4d) for up to 5% noise, after which the DE identified by PDE-FIND doesn't seem to model the dynamics. We compare the error of estimating the parameter values in the differential equation describing the simulated data. At 5% Gaussian noise the error of the fitting parameters increases to almost 80% (Figure 4b). The resulting integrated curve has MAE as low as 6 (on a color scale of 0-255) relative to the "ground truth" noise-free simulated curve (Figure 4c, Figure 4d). Additionally, PDE-FIND is no longer able to threshold sin and cosine terms to 0, as it even fits the noise with sinusoidal pattern.

We then consider different temperatures at the same noise level. The Verhulst logistic equation model becomes increasingly steep and shifts to the left with higher temperature. PDE-FIND successfully identifies this trend. It appears that the MAE is higher for equation extraction at higher-temperature data. This could be because of noise obscuring PDE-FIND's ability to fit steeper peaks accurately.

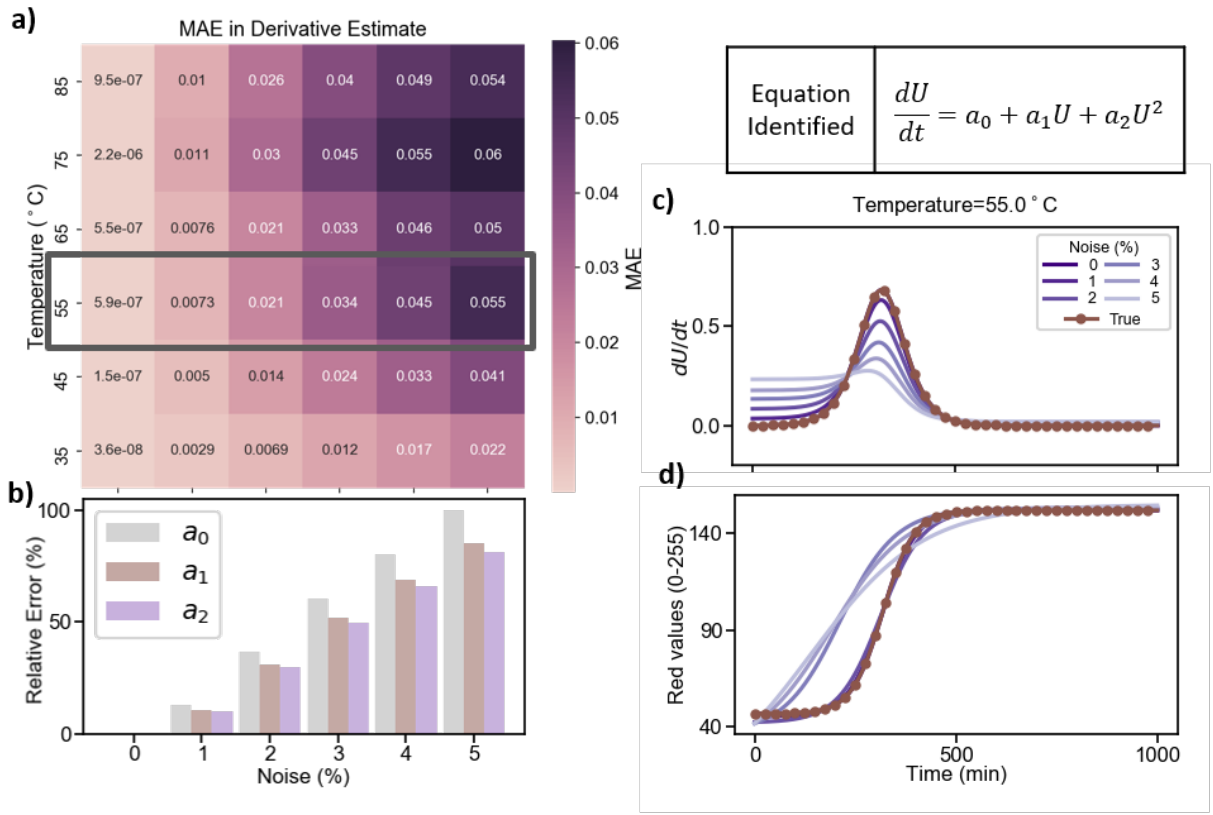

**Figure 4.** PDE-FIND Results: a) Heatmap of the mean absolute error (MAE) between the exact numerical derivative and the derivative estimate obtained from the DE identified by PDE-FIND at different noise levels at different temperatures for the 2$^{nd}$ order polynomial library. b) Relative error between the exact fitting parameter weight and those estimated by PDE-FIND at different noise levels for $T = 55$ °C data. c) Comparison of the exact numerical derivative at $T = 55$ °C with derivative estimated from the DE identified by PDE-FIND at different levels of noise. d)

Comparison of the exact solution at $T = 55$ °C with solution curves obtained by integrating the DE identified by PDE-FIND at different levels of noise.

**Discussion**

There remain many complex systems that have eluded quantitative analytic descriptions or even characterization of a suitable choice of variables in many disciplines such as biology, finance and materials science. With today's state-of-the art equipment, acquiring large quantities of data has never been easier. As put by Rackauckas *et al.*[59], "*the well-known adage 'a picture is worth a thousand words' might well be 'a model is worth a thousand datasets.'*".

   i.  *Scientific ML enables unique insights into MAPI degradation*

Scientific ML is a promising method that can be used to uncover governing equations through data, especially when the derivation of physical laws using first principles is challenging. In our study, we demonstrate that PDE-FIND identifies an underlying rate equation for the degradation of perovskite solar cells. MAPI degradation does not follow a simple single-order reaction rate law, defined as:

$$\frac{dU}{dt} = kU^n \tag{7}$$

where, *n* is the order of the reaction and *U* is the concentration of the species. In our system, this equation does not yield a good fit for *n=0, 1* or *2*. The S-shaped dynamics we see in our study have been reported in other studies involving MAPI degradation as well [46,49–51]. Some articles report that the degradation results from nucleation and growth of PbI$_2$ crystals[46,49], supporting the hypothesis that the kinetics follows the *Johnson-Mehl-Avrami-Kolmorgorov* or simply, the Avrami equation[47,48]:

$$\frac{dU'}{dt} = a_0 t^{n-1} - a_1 U' t^{n-1} \tag{8}$$

$$U'(t) = 1 - \exp(-kt^n) \tag{9}$$

Where,

$$U'(t) = \frac{U(t) - \min(U)}{\max(U - \min(U))}$$

And *a$_0$*, *a$_1$*, *n* and *k* are fitting constants.

Some recent studies have presented an alternate hypothesis of self-propagating or autocatalytic kinetics[50,51], which is described by another differential equation, the logistic function discussed earlier (Eqn 2, 3). In this study, we build a large library of candidate terms for the DE– polynomials of $U$, that make up the logistic function, and polynomials of $t$ and $U$ multiplied with polynomials of $t$, which feature in the Avrami equation. PDE-FIND determines that the simplest ODE that fits our experimental dataset best is of the form (Figure 3),

$$\frac{dU}{dt} = a_0 + a_1 U + a_2 U^2 \tag{6}$$

indicating the reaction is first, propelled forward by the presence of the reactant as well as the product, leading to a rapid growth in the product that eventually saturates when it exhausts its reactants – a self-propagating reaction. This is why we chose the logistic function model for the simulated dataset over the Avrami equations[41] that has been used to model nucleation-growth reactions. The algorithm picks terms that describe self-propelling kinetics (2nd order polynomial library) as opposed to diffusion-limited nucleation and growth (Avrami equation). When we examine videos of degrading films, we often observe a nucleation and growth behavior, whereby (lighter) regions of degraded material nucleate at specific points in the films, and expand with time. In the example shown in Figure S7 [and Supplementary Information video], one can see light areas of degraded material in the middle of the film degradation.

Equation (6) also offers insights that could help engineer more stable MAPI films. Once the degradation has begun, the autocatalytic nature suggests that degradation will continue, as the reaction products catalyze further MAPI degradation. Therefore, suppressing degradation means delaying the creation of the first reaction products for as long as possible. To engineer more stable MAPI films, this equation suggests that *reducing* MAPI degradation may be possible by reducing the density of nucleation points inside the material, including, *e.g.*, by ensuring that all $PbI_2$ precursors are fully converted during film formation, and possibly by using highly purified (i.e., devoid of contaminant particles) reagents in the film and adjacent layers that could nucleate $PbI_2$.

These insights bear consequence for researchers attempting to identify the underlying root cause(s) of perovskite degradation, as well as those modeling or predicting the (accelerated) degradation of these materials. If indeed this is a nucleation and growth phenomenon, little can be done to halt the growth of degraded regions once the initial nucleation event occurs. Therefore, to improve phase stability of perovskite films, an emphasis can be placed on identifying the nucleation points of these phase transformations, and inhibiting them, perhaps through improved precursor purification to remove impurities, improved control of the nucleation process, improved processing to remove growth catalysts, and improved packaging to prevent ingress of exogenous gasses. Changes to the film composition may increase the nucleation energy barrier; therefore, further investigation of stoichiometry optimization may be warranted in combination with the above.

*ii. Evaluating Scientific ML's ability to accommodate noisy experimental data*

We demonstrate the application of a scientific ML tool, PDE-FIND on MAPI degradation data. When applied to experimental data, PDE-FIND identifies a differential equation that fits the data, when appropriate constraints are applied. In spite of the noise and variance in the dataset, only functions corresponding to the dynamics of the system are picked and the DEs show good agreement with the numerical derivatives. Our "robustness analysis" with simulated data shows that PDE-FIND with a $2^{nd}$ order polynomial library succeeds at identifying the differential equation describing the simulated data when up to 5% Gaussian noise is added. However, the error of the fitting parameters increases with noise, to almost 80%. With 5% noise, the resulting integrated curve has a 6 MAE relative to the underlying noise-free simulated curve but the coefficients differ by as much as 80%. With the addition of noise, PDE-FIND is unable to eliminate terms not in the DE (sine and cosine) and even fits the noise with these terms.

Scientific ML methods can be immensely useful at uncovering governing equations of dynamical systems, if the data obtained has low noise or can be denoised by noise-reduction techniques. Data obtained through experiments is not devoid of measurement noise and denoising the data adequately can be challenging. Additionally, certain operating conditions cannot be fully controlled, leading to sample-to-sample variance making it hard to get rid of. Our contribution motivates the development of scientific ML techniques that are more robust to noise as well as variance in data. Scientific ML, in its current state, is well-suited to be applied to domains where obtaining large quantities of low-noise data is possible, and will find more applications with methods that are robust to noise.

### iii. Future opportunities for knowledge inference from experimental data

We show that Scientific ML has the potential to accelerate the understanding of materials degradation and the reliability optimization of perovskite materials. Extracting physical laws may facilitate the definition of acceleration factors for aging tests and also help in the prediction of perovskite solar cell degradation under varying environmental conditions. Not only does scientific machine learning aide us with understanding the underlying scientific phenomena better, it may also enable faster simulations and better extrapolations beyond our experimental datasets. The conclusions of any given materials study may well be rendered more generalizable by identifying underlying equations governing the observations.

**Acknowledgements**

The authors thank Kathleen Champion, Samuel Rudy, Zichao Long and Steven Atkinson for helpful discussions regarding scientific ML. This work was supported by Defense Advanced Research Projects Agency (DARPA) under contract no. HR001118C0036 (R.N.), TOTAL SA research grant funded through MITeI Sustng Mbr 9/08 (A.T., S.S., Z.L.), and the U.S. Department of Energy (DOE) under Photovoltaic Research and Development (PVRD) program under Award no. DE-EE0007535 (Z.L.). This work was partially supported by the U.S. Department of Energy's Office of Energy Efficiency and Renewable Energy (EERE) under the Advanced Manufacturing Office (AMO) Award Number DE-EE0009096 (R.N.). A.T. acknowledges the Alfred Kordelin Foundation.

**Author Contributions**

RN, AT, SS and TB conceived of and designed the study. CB and JT fabricated the samples. RN executed different aspects of the study such as the experiments and ML modelling. RN, AT and TB wrote the paper while all co-authors contributed to reviewing the manuscript.

**Conflict of Interest**

Although our laboratory has IP filed covering photovoltaic technologies and materials informatics broadly, we do not envision a direct COI with this study, the content of which is open sourced. Two of the authors (ZL, TB) own equity in a startup company applying machine learning to materials.